\documentclass[12pt]{article}

\def\be{\begin{equation}}
\def\ee{\end{equation}}

\newtheorem{theorem}{Theorem}
\newtheorem{lemma}[theorem]{Lemma}

\newtheorem{definition}[theorem]{Definition}

\begin{document}

\title{Replica symmetry breaking \\ related to a general ultrametric space I: \\
replica matrices and functionals}

\author{A.Yu.Khrennikov\footnote{International Center for Mathematical
Modelling in Physics and Cognitive Sciences, University of
V\"axj\"o, S-35195, Sweden, e--mail:
Andrei.Khrennikov@msi.vxu.se}, S.V.Kozyrev\footnote{Steklov
Mathematical Institute, Moscow, Russia, e--mail:
kozyrev@mi.ras.ru}}

\maketitle

\begin{abstract}
Family of replica matrices, related to general ultrametric spaces,
is introduced. These matrices generalize the known Parisi
matrices. Some functionals of replica approach are computed.
\end{abstract}

\section{Introduction}

One of the most interesting phenomenons of replica theory of spin
glasses and other disordered systems is the property of
ultrametricity of the replica space \cite{nature}, \cite{MPV}. In
papers \cite{ABK} and \cite{PaSu} it was shown that, in important
particular case, this ultrametric structure of replica space can
be described with the help of $p$--adic analysis, and the Parisi
replica matrix, which is the crucial ingredient of the replica
approach, can be considered as a $p$--adic pseudodifferential
operator. For other discussions on the relation of replicas with
analysis on ultrametric groups see also \cite{Carlucci1}. For
general introduction to the replica method see \cite{MPV}.

In papers \cite{ACHA}--\cite{nextIzv} a very general family of
ultrametric spaces was constructed, and a theory of ultrametric
pseudodifferential operators (or PDO) was developed. The bases of
ultrametric wavelets in spaces of functions on the constructed
ultrametric spaces were introduced, and it was shown that the
ultrametric PDO are diagonal in the bases of ultrametric wavelets.
This theory is a very far generalization of the results of the
papers \cite{wavelets}, \cite{nhoper}, where the corresponding
constructions for the $p$--adic case were introduced.

The mentioned above results are related to the field of $p$--adic
and ultrametric mathematical physics. For the other developments
in this field see \cite{VVZ}--\cite{Kochubei}. Among applications
of $p$--adic mathematical physics we would like to mention the
application to dynamics of macromolecules \cite{ABKO}.

In the present paper we, developing the approach of \cite{ABK} and
\cite{PaSu}, and propose a new replica symmetry breaking scheme,
based on a new family of replica block matrices, related to the
general family of ultrametric spaces and ultrametric PDO,
considered in \cite{ACHA}--\cite{Izv}.

We introduce a family of replica matrices, related to ultrametric
PDO of \cite{ACHA}--\cite{Izv}, and compute for these matrices
some functionals of the replica approach.

The organization of the present paper is as follows.

In Section 2 we describe parameterization of the Parisi matrices
by abelian groups with ultrametric.

In Section 3 we introduce the new family of replica block
matrices.

In Section 4 we show the relation between the introduced replica
matrices and restrictions of the ultrametric pseudodifferential
operators of \cite{ACHA}--\cite{Izv} on finite dimensional spaces
of test functions.

In Section 5 we compute some functionals of the replica approach
for introduced replica matrices.

In Section 6 we compare the results of computations of the
previous Section with the Parisi replica symmetry breaking scheme.

In Section 7 we put necessary material from
\cite{ACHA}--\cite{Izv} on ultrametric spaces and
pseudodifferential operators on these spaces.

\section{The Parisi matrices and their parameterization}

The Parisi replica matrix $\left(Q_{ab}\right)$ is the $n\times n$
matrix, defined as follows
$$
Q_{aa}=0
$$
\be\label{1.2} Q_{ab}=q_i \qquad \hbox{ if } \quad \left[{a\over
m_i}\right]\ne \left[{b\over m_i}\right],\quad \left[{a\over
m_{i+1}}\right]= \left[{b\over m_{i+1}}\right] \ee Here $m_i$ are
integers such that $p_i=m_{i+1}/m_i$ are integers,
$$
1=m_0< m_1<\dots< m_{k}< m_{k}=n
$$
and $q_i$ are $k$ real parameters, $[x]$ is the function of $x$,
which is equal to the smallest integer greater of equal to $x$.

In \cite{ABK}, \cite{PaSu}, it was shown that,  for the particular
case when $m_i=p^i$ (i.e. all $p_i=p$), formula (\ref{1.2}) is
equivalent to the following formula of the $p$--adic
parameterization: after the corresponding renumbering of the
indices of the Parisi replica matrix, the matrix elements of this
matrix can be put into the form \be\label{PM} Q_{ab}=q(|a-b|_p)
\ee where $|\cdot|_p$ is the $p$--adic distance and where the
function $q(x)$ encodes the information about the matrix elements
in the following way: $q(p^{i})=q_{i}$, $q(0)=0$.

Therefore the definition of this particular family of the Parisi
matrices is related to the simplest ultrametric spaces --- the
fields of $p$--adic numbers.

Let us generalize the construction of papers \cite{ABK},
\cite{PaSu} (of the $p$--adic parameterization of the Parisi
matrix) onto the case of general Parisi matrices (see also
\cite{Carlucci1} for the discussion of diagonalization of the
Parisi matrices by related construction). Consider the abelian
group (and ultrametric space), which is the direct product of
finite abelian cyclic groups: \be\label{groupG}
G=\prod_{i=1}^{k}{\bf Z}/p_i {\bf Z} \ee where ${\bf Z}$ is the
group of integers and $p_i$ are natural numbers. Elements of $G$
has the form $x=\{x_i\}=(x_1,\dots,x_k)$, $0\le x_i\le p_i-1$, and
the addition is elementwise modulo $p_i$:
$$
x+y=\{(x_i+y_i)\,{\rm mod }\, p_i\}
$$

\begin{lemma}{\sl
The elements of $G$ are in one to one correspondence with natural
numbers in $\{1,2,\dots,\prod_{j=1}^{k} p_j\}$. This
correspondence is given by the map
$$
l: 1,2,\dots,\prod_{j=1}^{k} p_j \to G
$$
\be\label{xinG} l^{-1}:\quad x=(x_1,\dots,x_k) \mapsto
1+\sum_{i=1}^{k} x_i \prod_{j=1}^{i-1} p_j ,\qquad 0\le x_i\le
p_i-1\ee Here we define $\prod_{j=1}^{0} p_j=1$.

}
\end{lemma}

\noindent{\it Proof}\qquad It is easy to see that the map $l^{-1}$
maps 0 in the group $G$ into 1. Also the image of
$$
(p_1-1,\dots,p_k-1)
$$
is
$$
1+\sum_{i=1}^{k} \left(p_i-1\right) \prod_{j=1}^{i-1}
p_j=1+\sum_{i=1}^{k} \left(\prod_{j=1}^{i} p_j-\prod_{j=1}^{i-1}
p_j\right)=\prod_{j=1}^{k} p_j
$$
The images of elements of the group $G$ will lie between 1 and
$\prod_{j=1}^{k} p_j$.

Check that the map $l^{-1}$ (which we call the enumeration map)
maps the different elements of the group $G$ into the different
numbers in $\{1,2,\dots,\prod_{j=1}^{k} p_j\}$.

Let $x\ne y$ be elements of the group $G$ such that $x_i\ne y_i$,
$x_j=y_j$ for $j>i$. Then
$$
l^{-1}(x)-l^{-1}(y)=\sum_{j=1}^{i} (x_j-y_j) \prod_{l=1}^{j-1}
p_l=\sum_{j=1}^{i-1} (x_j-y_j) \prod_{l=1}^{j-1} p_l+(x_i-y_i)
\prod_{l=1}^{i-1} p_l
$$
The module of the sum $\sum_{j=1}^{i-1} (x_j-y_j)
\prod_{l=1}^{j-1} p_l$ is not larger than $\prod_{l=1}^{i-1}
p_l-1$. The module of the contribution $(x_i-y_i)
\prod_{l=1}^{i-1} p_l$ is not less that $\prod_{l=1}^{i-1} p_l$,
thus the expression above is not equal to zero. This implies that
$l^{-1}$ maps the different elements of $G$ into the different
numbers. This finishes the proof of the lemma.

\bigskip

The group $G$ is ultrametric space with the ultrametric $|x|$
defined as
$$
|x|=m_{i}=\prod_{j=1}^{i} p_j
$$
where $i$ is the number of the last nonzero $x_i$.

We consider the Parisi matrix $Q=\left(Q_{ab}\right)$ as the
operator in the space of functions on the group $G$ with the
elements (\ref{xinG}). The following theorem takes place.

\begin{theorem}\label{p-adic_parameterization} {\qquad\sl The matrix element $Q_{ab}$, defined by
(\ref{1.2}), depends only on the ultrametric of the difference of
$l(a)$ and $l(b)$:
$$
Q_{ab}=q(|l(a)-l(b)|),
$$
where $q(m_i)=q_{i}$, $q(0)=0$, $i=1,\dots,k$. }
\end{theorem}

\noindent {\it Proof}\qquad The condition $\left[\frac{a} {m_{i}}
\right]= \left[\frac{b}{m_{i}} \right]$ takes the form
$$
\left[\frac{1+\sum_{i=1}^{k} a_i \prod_{j=1}^{i-1}
p_j}{m_{i}}\right]= \left[\frac{1+\sum_{i=1}^{k} b_i
\prod_{j=1}^{i-1} p_j}{m_{i}}\right]
$$
This implies that $a_j=b_j$ for $j> i$.

Analogously, the condition $\left[\frac{a} {m_{i}} \right]\ne
\left[\frac{b}{m_{i}} \right]$ means that $a_j\ne b_j$ for some
$j> i$.

The conditions $\left[\frac{a} {m_{i}} \right]=
\left[\frac{b}{m_{i}} \right]$, $\left[\frac{a}{m_{i-1}}\right]\ne
\left[\frac{b}{m_{i-1}}\right]$ taken together mean that $a_{i}\ne
b_{i}$, which implies $|l(a)-l(b)|=m_{i}$. We get that the matrix
element of the Parisi matrix $Q_{ab}$ depends only on the
ultrametric distance in $G$ of the indices $|l(a)-l(b)|$: if
$|l(a)-l(b)|$ equals to $m_{i}$, then $Q_{ab}=q_{i}$, which
implies the theorem.

\section{New family of replica block matrices}

In the present section we build a new family of replica matrices,
which generalize the Parisi matrices (\ref{PM}), used in the
replica approach \cite{physletta}, \cite{physrevlett}, \cite{MPV}.
These matrices will be some kind of block matrices, and each of
the constructed block matrices will correspond to some (directed)
tree.

We call a partially ordered set directed (and the corresponding
partial order --- a direction), if an arbitrary finite subset of
the partially ordered set has the unique supremum. We consider
directed trees, which define ultrametric spaces (see the Appendix
for details).

Consider a finite directed tree of the following form: we have the
vertex $I$, called the root of the tree. Then we have $p_I$, where
$p_I\ge 2$, vertices $I_j$, $j=0,\dots,p_I-1$, which are connected
by edges to the vertex $I$ and the partial order is defined as
$I_j<I$. Then, for (some, probably not for all) vertices $J=I_j$
we have $p_J$ vertices $J_j$, $j=0,\dots,p_J-1$, which are
connected by edges to the vertex $J$ and the partial order is
defined as $J_j<J$. Note that the branching indices $p_J$,
corresponding to different $J=I_j$ are not necessarily equal. Then
we iterate this procedure finite number of times (again, we may
take iteration not for all obtained vertices, but only for some
particular vertices), and obtain the finite directed tree ${\cal
S}$. We can obtain an arbitrary finite tree in this way (but not
an arbitrary finite {\it directed} tree).

Let us construct $n\times n$ block matrix $Q$, where $n$ is the
number of elements in the set ${\cal S}_{\,{\rm min}\,}$ of
minimal elements in the tree ${\cal S}$. Enumerate minimal
vertices $i\in {\cal S}_{\,{\rm min}\,}$ by $i=1,\dots,n$. This
index will enumerate the rows and columns of the block matrix $Q$.
To each of the vertices $J\in {\cal S}\backslash {\cal S}_{\,{\rm
min}\,}$ we put into correspondence a number $q_J$.

Then we define the block matrix $Q$, for which the diagonal matrix
elements are zero:
$$ Q_{ii}=0;$$
and the matrix element $Q_{ij}$, $i\ne j$, $i,j=1,\dots,n$, is
defined using the partial order in the tree ${\cal S}$:
\be\label{sup_formula} Q_{ij}=q_{\,{\rm
sup}\,(i,j)}\sqrt{\mu_i\mu_j}, \qquad i\ne j \ee where $\,{\rm
sup}\,(i,j)$ is the vertex in ${\cal S}$ which is the supremum
with respect to the partial order in ${\cal S}$ of the minimal
vertices with the numbers $i$ and $j$ (the supremum of two
vertices is the smallest vertex, which is larger both of the
vertices). Here $q_{\,{\rm sup}\,(i,j)}$ is the mentioned above
number, which corresponds to the vertex ${\,{\rm sup}\,(i,j)}$,
and $\mu_i$, $\mu_j$ are some positive numbers (equal to measures
of balls in some ultrametric space), see the definition in the
next section.

Formula (\ref{sup_formula}), which allows to describe the
introduced replica block matrices in terms of directed trees is
the crucial ingredient in the approach of the present paper. This
formula can be considered as a natural generalization of $p$--adic
parameterization of the Parisi matrix (\ref{PM}) onto the case of
general ultrametric spaces.

Consider examples of the introduced replica matrices.

\bigskip

\noindent{\bf Example 1}\qquad The simplest example of the matrix
of the described in the present section form (which differs from
the Parisi matrix) is \be\label{example} Q_1=\pmatrix{0 & q_1 &
q_0 & q_0 & q_0 \cr q_1 & 0 & q_0 & q_0 & q_0\cr q_0 & q_0 & 0 &
q_2 & q_2 \cr q_0 & q_0 & q_2 & 0 & q_2 \cr q_0 & q_0 & q_2 & q_2
& 0\cr } \ee Here we have two blocks, of the sizes $2\times 2$ and
$3\times 3$, which are combined in the $2\times 2$ block matrix
with the described blocks at the diagonal. The resulting dimension
of the matrix is $5\times 5$.

By the procedure, described at the present section, it corresponds
to the directed tree with the vertices, enumerated as  0, 1, 2,
10, 11, 20, 21, 22. The partial order on the vertices is defined
as follows:
$$
0>1,\quad 0>2,\quad 1>10,\quad 1>11,\quad 2>20,\quad 2>21,\quad
2>22
$$
The pairs of ordered vertices listed above are connected by
(directed) edges.

The vertices of the tree correspond to the matrix elements of the
matrix as follows. The vertex 0 correspond to the value $q_0$, the
vertex 1 correspond to the value $q_1$, the vertex 2 correspond to
the value $q_2$. The lowest vertices 10, 11, 20, 21, 22 enumerate
the rows and the columns of the matrix (in the above order). The
vertices 1 and 2 enumerate the blocks, and one can say that the
vertex 0 enumerate the matrix itself.

\bigskip

\noindent{\bf Example 2}\qquad The second more complicated example
of the block matrix  is  \be\label{example1} Q_2=\pmatrix{ 0 & q_3
& q_1 & q_1 & q_1 & q_0 & q_0 \cr q_3 & 0 & q_1 & q_1 & q_1 & q_0
& q_0 \cr q_1 & q_1 & 0 & q_4 & q_4 & q_0 & q_0 \cr q_1 & q_1 &
q_4 & 0 & q_4 & q_0 & q_0 \cr q_1 & q_1 & q_4 & q_4 & 0 & q_0 &
q_0 \cr q_0 & q_0 & q_0 & q_0 & q_0 & 0 & q_2 \cr q_0 & q_0 & q_0
& q_0 & q_0 & q_2 & 0\cr} \ee

Here we have two blocks, of the sizes $2\times 2$ and $5\times 5$,
which are combined in the $7\times 7$ block matrix with the
mentioned blocks at the diagonal. The $5\times 5$ block is the
block matrix $Q_1$ from the previous example (which itself
contains nontrivial blocks).

\section{Ultrametric PDO and replica matrices}

In the present section we show that the introduced block matrices
are related to ultrametric pseudodifferential operators (PDO) on
ultrametric spaces, considered in \cite{ACHA}, \cite{Izv}, see the
Appendix for discussion. The construction of ultrametric spaces
and the corresponding PDO is based on a directed tree, which we
will denote ${\cal T}$. Consider a subtree ${\cal S}\subset {\cal
T}$ satisfying the following definition.

\begin{definition}\label{wave_type}
{\sl The subset ${\cal S}$ in a directed tree ${\cal T}$ (with the
partial order of the kind considered in the Appendix) is called of
the analytic type, iff:

1) ${\cal S}$ is finite;

2) ${\cal S}$ is a directed subtree in ${\cal T}$ (where the
direction in ${\cal S}$ is the restriction of the direction in
${\cal T}$ onto ${\cal S}$);

3) The directed subtree ${\cal S}$ obey the following property: if
${\cal S}$ contains a vertex $I$ and a vertex $J$: $J<I$,
$|IJ|=1$, then the subtree ${\cal S}$ contains all the vertices
$L$ in ${\cal T}$: $L<I$, $|IL|=1$. }
\end{definition}

The maximal vertex in ${\cal S}$ we will denote $K$.

Absolute of the directed tree ${\cal T}$, see \cite{ACHA},
\cite{Izv} and the Appendix for the details, is the ultrametric
space $X({\cal T})$ with the naturally defined measure $\mu$.
Vertices of the tree ${\cal T}$ are in one to one correspondence
with balls (or disks) in $X({\cal T})$. We denote by $J$ the disk
in $X({\cal T})$, corresponding to vertex $J$, and by $\chi_J$ we
denote the characteristic function of this disk.

For the finite subtree ${\cal S}\subset {\cal T}$ of the analytic
type  consider the space $D({\cal S})$ \footnote{The space
$D({\cal S})$ is a finite dimensional space of test functions on
the ultrametric space $X$ and is an analogue of the
Bruhat--Schwartz space of test functions of $p$--adic argument.
The space $D(X)$ of test functions on the absolute is the
inductive limit of the spaces $D({\cal S})$: $D(X)=\lim\,{\rm
ind}_{{\cal S}\to {\cal T}}\, D({\cal S})$.}, which is the linear
span of vectors $\chi_J$ with $J\in {\cal S}$. We consider this
space as the subspace in the space $L^2(X,\mu)$ of quadratically
integrable with respect to the measure $\mu$ functions on the
absolute. Obviously in the space $D({\cal S})$ there is the
orthonormal basis $\left\{{\chi_J\over \sqrt{\mu(J)}}\right\}$
with $J$ running over the minimal vertices in ${\cal S}$, i.e. a
vector $f$ in $D({\cal S})$ can be put into the form
\be\label{funinS} f=\sum_{J\in {\cal S}_{\rm min}}f_J {\chi_J\over
\sqrt{\mu(J)}} \ee where the summation runs over minimal elements
in ${\cal S}$.

Consider the operator $Q({\cal S})=\Pi({\cal S})T$ in $D({\cal
S})$, where $\Pi({\cal S})$ is the orthogonal projection in the
space $L^2(X,\mu)$ onto $D({\cal S})$, and $T$ is an ultrametric
pseudodifferential operator (PDO), see the Appendix. Operator
$Q({\cal S})$ is an operator in finite dimensional linear space
$D({\cal S})$.

\begin{lemma}\label{theformula}{\sl
Action of $Q({\cal S})=\Pi({\cal S})T$ on the functions
(\ref{funinS}) in $D({\cal S})$ is given by the formula
$$
(Q({\cal S})f)_I=\sum_{J\in {\cal S}_{\rm
min}}\sqrt{\mu(I)\mu(J)}T{(\,{\rm
sup}\,(I,J))}\left[\sqrt{\mu(J)\over\mu(I)}f_I -
f_J\right]+C_{\cal S}f_I;
$$
where
$$
C_{\cal S}=\int_{y\notin K }
d\mu(y)T(I,y)=\sum_{L>K}T{(L)}\mu(L)\left(1-p_L\right)^{-1}
$$
and $I$, $J$ run over the minimal vertices in ${\cal S}$. }
\end{lemma}

\noindent{\it Proof}\qquad Take $I\in {\cal S}_{\rm min}$ and
consider the scalar product in $L^2(X,\mu)$:
$$
\left\langle {\chi_I\over \sqrt{\mu(I)}}, T\sum_{J\in {\cal
S}_{\rm min}}f_J {\chi_J\over \sqrt{\mu(J)}}\right\rangle=$$
$$
=\sum_{J\in {\cal S}_{\rm min}}{1\over
\sqrt{\mu(I)\mu(J)}}f_J\int_{x\in I}d\mu(x)\int
d\mu(y)T(x,y)(\chi_J(x)-\chi_J(y))=
$$
$$
={1\over \mu(I)}f_I\int_{x\in I}d\mu(x)\int
d\mu(y)T(x,y)(\chi_I(x)-\chi_I(y))+
$$
$$
+\sum_{J\in {\cal S}_{\rm min},J\ne I}{1\over
\sqrt{\mu(I)\mu(J)}}f_J\int_{x\in I}d\mu(x)\int
d\mu(y)T(x,y)(\chi_J(x)-\chi_J(y))=
$$
$$
={1\over \mu(I)}f_I\int_{x\in I}d\mu(x)\int_{y\notin I }
d\mu(y)T(x,y)-
$$
$$
-\sum_{J\in {\cal S}_{\rm min},J\ne I}{1\over
\sqrt{\mu(I)\mu(J)}}T{(\,{\rm sup}\,(I,J))}f_J\int_{x\in
I}d\mu(x)\int d\mu(y)\chi_J(y)=
$$
$$
=f_I\int_{y\notin I } d\mu(y)T(I,y)- \sum_{J\in {\cal S}_{\rm
min},J\ne I}\sqrt{\mu(I)\mu(J)}T{(\,{\rm sup}\,(I,J))}f_J=
$$
$$
=f_I\int_{y\notin K } d\mu(y)T(I,y)+f_I\sum_{J\in {\cal S}_{\rm
min},J\ne I}\mu(J)T{(\,{\rm sup}\,(I,J))}-
$$
$$
-\sum_{J\in {\cal S}_{\rm min},J\ne I}\sqrt{\mu(I)\mu(J)}T{(\,{\rm
sup}\,(I,J))}f_J=
$$
$$
=f_I\int_{y\notin K } d\mu(y)T(I,y)+\sum_{J\in {\cal S}_{\rm
min}}\sqrt{\mu(I)\mu(J)}T{(\,{\rm
sup}\,(I,J))}\left[\sqrt{\mu(J)\over\mu(I)}f_I - f_J\right].
$$
This proves the lemma.

\bigskip

Therefore, the operator $Q({\cal S})$ corresponds to the block
matrix (\ref{sup_formula}): \be\label{Q}
Q_{IJ}=\sqrt{\mu(I)\mu(J)}T{(\,{\rm sup}\,(I,J))},\qquad I,J\in
{\cal S}_{\rm min} \ee defined at the previous section. The lemma
above shows that this matrix is related to the restriction of
action of the ultrametric PDO onto the space $D({\cal S})$,
corresponding to the subtree ${S}\subset {\cal T}$ of the analytic
type. In the following we will omit the index ${\cal S}$ in the
notation of the replica matrix $Q({\cal S})$, i.e. we will write
$Q$ instead of $Q({\cal S})$.

\bigskip

\noindent{\bf Remark}\qquad It is easy to construct the
generalization of the Parisi anzats on the case of general
ultrametric spaces. The generalization of the Parisi matrix will
look like \be\label{QP}
Q_{IJ}=\sqrt{\mu(I)\mu(J)}F\left(\mu({(\,{\rm
sup}\,(I,J))})\right),\quad I,J\in {\cal S}_{\rm min}, \quad
F(x)\ge 0, \ee where $F$ is a non--negative function of real
argument. Let us note that the matrix (\ref{Q}) is more general
than the matrix (\ref{QP}), even in the $p$--adic case.

\bigskip

\noindent{\bf Remark}\qquad Note that $T(I,y)$ in the expression
for $C_{\cal S}$ does not depend on $I\in {\cal S}$ and depends
only on $K={\rm sup}\, {\cal S}$. The constant $C_{\cal S}$ tends
to zero if the largest vertex of ${\cal S}$ tends to infinity. We
will ignore the contribution containing $C_{\cal S}$ in the
following.

\section{Computations with replica matrices}

In the present section we compute the products and traces of the
degrees for the introduced in the previous section matrices. Some
similar computations for $p$--adic case can be found in
\cite{PaSu}. The next lemma gives some useful formulas for
computations, related to subtrees of analytic type.

\begin{lemma}\label{formulae}
{\sl
$$
\sum_{J\in {\cal S}_{\rm min}:\,{\rm
sup}\,(I,J)=L}\mu(J)=\mu(L)\left(1-p_L^{-1}\right)
$$
$$
\sum_{I,J\in {\cal S}_{\rm min}:\,{\rm
sup}\,(I,J)=L}\mu(I)\mu(J)=\mu^2(L)\left(1-p_L^{-1}\right)
$$
}
\end{lemma}

Proof of the lemma is by additivity of the measure $\mu$.

The following lemma computes the important in the replica approach
functional ${\rm tr}\,Q=\sum_{a\ne b}Q_{ab}$.

\begin{lemma}\label{traceRMst}
{\sl
$$
\sum_{I,J\in {\cal S}_{\rm min};I\ne J}\mu(I){\mu(J)}{T{(\,{\rm
sup}\,(I,J))}}= \sum_{J\in {\cal S}\backslash {\cal S}_{\rm
min}}T{(J)}\mu^2(J)\left(1-p_J^{-1}\right)
$$
}
\end{lemma}

\noindent{\it Proof}\qquad Computing the sum over $J$ we get
$$
\sum_{I,J\in {\cal S}_{\rm min};I\ne J}\mu(I){\mu(J)}{T{(\,{\rm
sup}\,(I,J))}}= \sum_{I\in {\cal S}_{\rm min}}\mu(I)\sum_{I< L\le
K}{T{(L)}}\sum_{J\in {\cal S}_{\rm min}:\,{\rm
sup}\,(I,J)=L}{\mu(J)}
$$
where $K$ is the largest vertex in ${\cal S}$. This reduces to
$$
\sum_{I\in {\cal S}_{\rm min}}\mu(I)\sum_{I< L\le
K}{T{(L)}}\mu(L)\left(1-p_L^{-1}\right)=\sum_{J\in {\cal
S}\backslash {\cal S}_{\rm
min}}{T{(J)}}\mu(J)\left(1-p_J^{-1}\right)\sum_{I\in {\cal S}_{\rm
min}:I<J}\mu(I)=
$$
$$
=\sum_{J\in {\cal S}\backslash {\cal S}_{\rm
min}}{T{(J)}}\mu^2(J)\left(1-p_J^{-1}\right)
$$
This finishes the proof of the lemma.

\bigskip

Prove the following important lemma, which allows to compute
products of replica matrices in terms of summation over the paths
in corresponding trees.

\begin{lemma}\label{prod_mat}{\sl
For $I,J\in {\cal S}_{\rm min}$ one has: if $I\ne J$, then
$$
\sum_{L\in {\cal S}_{\rm min},L\ne I,J}\mu(L)T_1{(\,{\rm
sup}\,(I,L))}T_2{(\,{\rm sup}\,(L,J))} =
$$
$$
=\mu\left({\,{\rm sup}\,(I,J)}\right)\left(1-2p^{-1}_{\,{\rm
sup}\,(I,J)}\right)T_1{(\,{\rm sup}\,(I,J))}T_2{(\,{\rm
sup}\,(I,J))}+
$$
$$
+\sum_{L:\,{\rm sup}\,(I,J)< L\le
K}\mu(L)\left(1-p_L^{-1}\right)T_1{(L)}T_2{(L)} +
$$
$$
+\left[\sum_{L:I<L<\,{\rm
sup}\,(I,J)}\mu(L)\left(1-p_L^{-1}\right)T_1{(L)}\right]T_2{(\,{\rm
sup}\,(I,J))}+
$$
$$
+T_1{(\,{\rm sup}\,(I,J))}\left[\sum_{L:J<L<\,{\rm
sup}\,(I,J)}\mu(L)\left(1-p_L^{-1}\right)T_2{(L)}\right]
$$
and for $I=J$ one has
$$
\sum_{L\in {\cal S}_{\rm min},L\ne I}\mu(L)T_1{(\,{\rm
sup}\,(I,L))}T_2{(\,{\rm sup}\,(L,I))} =\sum_{L:I<L\le
K}\mu(L)\left(1-p_L^{-1}\right)T_1{(L)}T_2{(L)}
$$
In particular \be\label{sum_T_IJ} \sum_{I,J\in {\cal S}_{\rm
min};I\ne J}\mu(I){\mu(J)}{T{(\,{\rm sup}\,(I,J))}}^2=\sum_{J\in
{\cal S}\backslash {\cal S}_{\rm
min}}{T{(J)}}^2\mu^2(J)\left(1-p_J^{-1}\right) \ee

}
\end{lemma}

\noindent{\it Proof}\qquad Compute for $I\ne J$
$$
\sum_{L\in {\cal S}_{\rm min},L\ne I,J}\mu(L)T_1{(\,{\rm
sup}\,(I,L))}T_2{(\,{\rm sup}\,(L,J))} =
$$
$$
=\sum_{L\in {\cal S}_{\rm min},L\ne I,J:\,{\rm sup}\,(L,\,{\rm
sup}\,(I,J))> \,{\rm sup}\,(I,J) }\mu(L)T_1{(\,{\rm
sup}\,(I,L))}T_2{(\,{\rm sup}\,(L,J))}+
$$
$$
+\sum_{L\in {\cal S}_{\rm min},L\ne I,J:\,{\rm sup}\,(L,I)<\,{\rm
sup}\,(I,J)}\mu(L)T_1{(\,{\rm sup}\,(I,L))}T_2{(\,{\rm
sup}\,(L,J))}+
$$
$$
+\sum_{L\in {\cal S}_{\rm min},L\ne I,J:\,{\rm sup}\,(L,J)<\,{\rm
sup}\,(I,J)}\mu(L)T_1{(\,{\rm sup}\,(I,L))}T_2{(\,{\rm
sup}\,(L,J))}+
$$
$$
+\sum_{L\in {\cal S}_{\rm min},L\ne I,J:\,{\rm sup}\,(L,I)=\,{\rm
sup}\,(L,J)=\,{\rm sup}\,(I,J)}\mu(L)T_1{(\,{\rm
sup}\,(I,L))}T_2{(\,{\rm sup}\,(L,J))}
$$
$$
=\sum_{L:\,{\rm sup}\,(I,J)<L\le
K}\mu(L)\left(1-p_L^{-1}\right)T_1{(L)}T_2{(L)}+
$$
$$
+\sum_{L:I<L<\,{\rm
sup}\,(I,J)}\mu(L)\left(1-p_L^{-1}\right)T_1{(L)}T_2{(\,{\rm
sup}\,(I,J))}+
$$
$$
+\sum_{L:J<L<\,{\rm
sup}\,(I,J)}\mu(L)\left(1-p_L^{-1}\right)T_1{(\,{\rm
sup}\,(I,J))}T_2{(L)}+
$$
$$
+ \mu\left({\,{\rm sup}\,(I,J)}\right)\left(1-2p^{-1}_{\,{\rm
sup}\,(I,J)}\right)T_1{(\,{\rm sup}\,(I,J))}T_2{(\,{\rm
sup}\,(I,J))}
$$
Analogously, for $I=J$ one obtains
$$
\sum_{L\in {\cal S}_{\rm min},L\ne I}\mu(L)T_1{(\,{\rm
sup}\,(I,L))}T_2{(\,{\rm sup}\,(L,I))} =\sum_{L:I<L\le
K}\mu(L)\left(1-p_L^{-1}\right)T_1{(L)}T_2{(L)}
$$
This finishes the proof of the lemma.

\bigskip

The next lemma gives the trace of the cubic combination of the
replica matrices.

\begin{lemma}\label{trQ3}{\sl For the trace of the cubic combination of the
replica block matrices we get
$$
\,{\rm tr}\,Q_1Q_2Q_3=
$$
$$
=\sum_{A,B,C\in {\cal S}_{\rm min},A\ne
B,B\ne C,C\ne A}\mu(A)\mu(B)\mu(C)T_1{(\,{\rm
sup}\,(A,B))}T_2{(\,{\rm sup}\,(B,C))}T_3{(\,{\rm sup}\,(C,A))} =
$$
$$
=\sum_{L\in {\cal S}\backslash {\cal S}_{\rm
min}}\mu^3(L)\left(1-p_L^{-1}\right)\left(1-2p^{-1}_{L}\right)T_1{(L)}T_2{(L)}T_3{(L)}+
$$
$$
+\sum_{L\in {\cal S}\backslash {\cal S}_{\rm
min}}T_1{(L)}T_2{(L)}\mu(L)\left(1-p_L^{-1}\right)
\left[\sum_{B\in {\cal S}\backslash {\cal S}_{\rm min}
:B<L}\mu^2(B)\left(1-p_B^{-1}\right)T_3{(B)}\right]+
$$
$$
+\sum_{L\in {\cal S}\backslash {\cal S}_{\rm
min}}T_2{(L)}T_3{(L)}\mu(L)\left(1-p_L^{-1}\right)
\left[\sum_{B\in {\cal S}\backslash {\cal S}_{\rm min}
:B<L}\mu^2(B)\left(1-p_B^{-1}\right)T_1{(B)}\right]+
$$
$$
+\sum_{L\in {\cal S}\backslash {\cal S}_{\rm
min}}T_3{(L)}T_1{(L)}\mu(L)\left(1-p_L^{-1}\right)
\left[\sum_{B\in {\cal S}\backslash {\cal S}_{\rm min}
:B<L}\mu^2(B)\left(1-p_B^{-1}\right)T_2{(B)}\right]
$$
}
\end{lemma}

\noindent{\bf Remark}\qquad In more compact form the above formula
can be written as \be\label{ultra} \sum_{A,B,C\in {\cal S}_{\rm
min},A\ne B,B\ne C,C\ne A}\mu(A)\mu(B)\mu(C)T_1{(\,{\rm
sup}\,(A,B))}T_2{(\,{\rm sup}\,(B,C))}T_3{(\,{\rm sup}\,(C,A))} =
$$
$$
=\sum_{L\in {\cal S}\backslash {\cal S}_{\rm
min}}\mu^3(L)\left(1-p_L^{-1}\right)\left(1-2p^{-1}_{L}\right)T_1{(L)}T_2{(L)}T_3{(L)}+
$$
$$
+\sum_{L\in {\cal S}\backslash {\cal S}_{\rm
min}}\mu(L)\left(1-p_L^{-1}\right) \sum_{B\in {\cal S}\backslash
{\cal S}_{\rm min} :B<L}\mu^2(B)\left(1-p_B^{-1}\right)
$$
$$
\left[T_1{(L)}T_2{(L)}T_3{(B)}+T_2{(L)}T_3{(L)}T_1{(B)}+T_3{(L)}T_1{(L)}T_2{(B)}\right]
\ee The property of the sum in (\ref{ultra}), that two of the
three (or even three of the three) indices $I$ of the coefficients
$T{(I)}$ coincide, and the two coinciding indices are larger than
the third, is related to the fact that in ultrametric space all
triangles are equilateral with equal larger edges. The analogous
observations for the Parisi RSB ansatz were made in \cite{nature}.
Note that here (unlike in \cite{nature}) we did not yet introduce
any kind of the $n\to 0$ limit.

\bigskip

\noindent{\it Proof of Lemma \ref{trQ3}}\qquad Compute
$$
\sum_{A,B,C\in {\cal S}_{\rm min},A\ne B,B\ne C,C\ne
A}\mu(A)\mu(B)\mu(C)T_1{(\,{\rm sup}\,(A,B))}T_2{(\,{\rm
sup}\,(B,C))}T_3{(\,{\rm sup}\,(C,A))} =
$$
$$
=\sum_{A,C\in {\cal S}_{\rm min};C\ne
A}\mu(A)\Biggl[\sum_{B:\,{\rm sup}\,(A,C)< B\le
K}\mu(B)\left(1-p_B^{-1}\right)T_1{(B)}T_2{(B)} +
$$
$$
+ \mu\left({\,{\rm sup}\,(A,C)}\right)\left(1-2p^{-1}_{\,{\rm
sup}\,(A,C)}\right)T_1{(\,{\rm sup}\,(A,C))}T_2{(\,{\rm
sup}\,(A,C))}+
$$
$$
+\left[\sum_{B:A<B<\,{\rm
sup}\,(A,C)}\mu(B)\left(1-p_B^{-1}\right)T_1{(B)}\right]T_2{(\,{\rm
sup}\,(A,C))}+
$$
$$
+T_1{(\,{\rm sup}\,(A,C))}\left[\sum_{B:C<B<\,{\rm
sup}\,(A,C)}\mu(B)\left(1-p_B^{-1}\right)T_2{(B)}\right]\Biggr]\mu(C)T_3{(\,{\rm
sup}\,(C,A))}
$$
Discuss the third contribution to the above sum, i.e. the
expression
$$
\sum_{A,C\in {\cal S}_{\rm min};C\ne A}\mu(A)\mu(C)T_2{(\,{\rm
sup}\,(A,C))}T_3{(\,{\rm sup}\,(C,A))}\left[\sum_{B:A<B<\,{\rm
sup}\,(A,C)}\mu(B)\left(1-p_B^{-1}\right)T_1{(B)}\right]
$$
In this expression we put the summation as the following
composition of the three summations
$$
\sum_{A,C\in {\cal S}_{\rm min};C\ne A}=\sum_{C\in {\cal S}_{\rm
min}}\sum_{L:C<L\le K}\sum_{A\in {\cal S}_{\rm min}:\,{\rm
sup}\,(A,C)=L}
$$
This implies
$$
\sum_{C\in {\cal S}_{\rm min}}\sum_{L:C<L\le K}\sum_{A\in {\cal
S}_{\rm min}:\,{\rm sup}\,(A,C)=L}\mu(A)\mu(C)T_2{(\,{\rm
sup}\,(A,C))}T_3{(\,{\rm sup}\,(C,A))}$$
$$
\left[\sum_{B:A<B<\,{\rm
sup}\,(A,C)}\mu(B)\left(1-p_B^{-1}\right)T_1{(B)}\right]=
$$
$$
=\sum_{C\in {\cal S}_{\rm min}}\mu(C)\sum_{L:C<L\le
K}T_2{(L)}T_3{(L)} \sum_{A\in {\cal S}_{\rm min}:\,{\rm
sup}\,(A,C)=L}\mu(A)\left[\sum_{B:A<B<L}\mu(B)\left(1-p_B^{-1}\right)T_1{(B)}\right]=
$$
$$
=\sum_{C\in {\cal S}_{\rm min}}\mu(C)\sum_{L:C<L\le
K}T_2{(L)}T_3{(L)} \left[\sum_{B\in {\cal S}\backslash {\cal
S}_{\rm min} :B<L,\,{\rm
sup}\,(B,C)=L}\mu^2(B)\left(1-p_B^{-1}\right)T_1{(B)}\right]=
$$
$$
=\sum_{L\in {\cal S}\backslash {\cal S}_{\rm
min}}T_2{(L)}T_3{(L)}\sum_{C\in {\cal S}_{\rm min}:C<L}\mu(C)
\left[\sum_{B\in {\cal S}\backslash {\cal S}_{\rm min} :B<L,\,{\rm
sup}\,(B,C)=L}\mu^2(B)\left(1-p_B^{-1}\right)T_1{(B)}\right]=
$$
\be\label{the3rd} =\sum_{L\in {\cal S}\backslash {\cal S}_{\rm
min}}T_2{(L)}T_3{(L)}\mu(L)\left(1-p_L^{-1}\right)
\left[\sum_{B\in {\cal S}\backslash {\cal S}_{\rm min}
:B<L}\mu^2(B)\left(1-p_B^{-1}\right)T_1{(B)}\right] \ee
Analogously, the fourth contribution
$$
\sum_{A,C\in {\cal S}_{\rm min};C\ne A}\mu(A)\mu(C)T_1{(\,{\rm
sup}\,(A,C))}T_3{(\,{\rm sup}\,(C,A))}\left[\sum_{B:C<B<\,{\rm
sup}\,(A,C)}\mu(B)\left(1-p_B^{-1}\right)T_2{(B)}\right]
$$
will give \be\label{the4th} \sum_{L\in {\cal S}\backslash {\cal
S}_{\rm min}}T_1{(L)}T_3{(L)}\mu(L)\left(1-p_L^{-1}\right)
\left[\sum_{B\in {\cal S}\backslash {\cal S}_{\rm min}
:B<L}\mu^2(B)\left(1-p_B^{-1}\right)T_2{(B)}\right] \ee Compute
the first contribution. We get
$$
\sum_{A,C\in {\cal S}_{\rm min};C\ne A}\mu(A)\mu(C)T_3{(\,{\rm
sup}\,(C,A))}\left[\sum_{B:\,{\rm sup}\,(A,C)< B\le
K}\mu(B)\left(1-p_B^{-1}\right)T_1{(B)}T_2{(B)} \right]=
$$
$$
=\sum_{B\in {\cal S}\backslash {\cal S}_{\rm
min}}\mu(B)\left(1-p_B^{-1}\right)T_1{(B)}T_2{(B)}
\left[\sum_{A,C\in {\cal S}_{\rm min};C\ne
A;A,C<B}\mu(A)\mu(C)T_3{(\,{\rm sup}\,(C,A))}\right]=
$$
\be\label{the1st} =\sum_{B\in {\cal S}\backslash {\cal S}_{\rm
min}}\mu(B)\left(1-p_B^{-1}\right)T_1{(B)}T_2{(B)}
\left[\sum_{L\in {\cal S}\backslash {\cal S}_{\rm min}
:L<B}\mu^2(L)\left(1-p_L^{-1}\right)T_3{(L)}\right] \ee The second
contribution reduces to
$$
\sum_{A,C\in {\cal S}_{\rm min};C\ne A}\mu(A)\mu(C)
\mu\left({\,{\rm sup}\,(A,C)}\right)\left(1-2p^{-1}_{\,{\rm
sup}\,(A,C)}\right)
$$ $$
T_1{(\,{\rm sup}\,(A,C))}T_2{(\,{\rm
sup}\,(A,C))}T_3{(\,{\rm sup}\,(C,A))}=
$$
\be\label{the2nd} =\sum_{L\in {\cal S}\backslash {\cal S}_{\rm
min}}\mu^3(L)\left(1-p_L^{-1}\right)\left(1-2p^{-1}_{L}\right)T_1{(L)}T_2{(L)}T_3{(L)}
\ee Combining formulas (\ref{the3rd})--(\ref{the2nd}), we obtain
the proof of the lemma.

\bigskip

The next lemma gives the trace of the quartic combination.

\begin{lemma}\label{trQ4}{\sl
$$
\,{\rm tr}\,Q^4=\sum_{A,B,C,D\in {\cal S}_{\rm min},A\ne B,B\ne
C,C\ne D,D\ne A}\mu(A)\mu(B)\mu(C)\mu(D)
$$
$$
T{(\,{\rm sup}\,(A,B))}T{(\,{\rm sup}\,(B,C))}T{(\,{\rm
sup}\,(C,D))}T{(\,{\rm sup}\,(D,A))} =
$$
$$
=\sum_{L\in {\cal S}\backslash {\cal S}_{\rm
min}}\mu^4\left({L}\right)\left(1-p^{-1}_{L}\right)\left(1-2p^{-1}_{L}\right)^2{T{(L)}}^4+
$$
$$
+2\sum_{L\in {\cal S}\backslash {\cal S}_{\rm
min}}\mu^3\left({L}\right)\left(1-p^{-1}_{L}\right)\left(1-2p^{-1}_{L}\right){T{(L)}}^2
\sum_{M:L< M\le K}\mu(M)\left(1-p_M^{-1}\right){T{(M)}}^2 +
$$
$$
+4\sum_{L\in {\cal S}\backslash {\cal S}_{\rm min}}
\mu\left({L}\right)\left(1-2p^{-1}_{L}\right){T{(L)}}^3
\left[\sum_{M\in {\cal S}\backslash {\cal S}_{\rm min}
:M<L}\mu^3(M)\left(1-p_M^{-1}\right)^2T{(M)}\right]+
$$
$$
+\sum_{M\in {\cal S}\backslash {\cal S}_{\rm
min}}\mu^2\left({M}\right)\left(1-p^{-1}_{M}\right)\left[\sum_{L:M<
L\le K}\mu(L)\left(1-p_L^{-1}\right){T{(L)}}^2\right]^2+
$$
$$
+4\sum_{I\in {\cal S}\backslash {\cal S}_{\rm
min}}\left[\sum_{M\in {\cal S}\backslash {\cal S}_{\rm
min}:M<I}\mu^3(M)\left(1-p_M^{-1}\right)^2T{(M)}\right]T{(I)}\left[\sum_{L:I<
L\le K}\mu(L)\left(1-p_L^{-1}\right){T{(L)}}^2\right]
$$
$$
+2\sum_{L\in {\cal S}\backslash {\cal S}_{\rm
min}}\left[\sum_{M\in {\cal S}\backslash {\cal S}_{\rm
min}:M<L}\mu^4(M)\left(1-p_M^{-1}\right)^3{T{(M)}}^2\right]{T{(L)}}^2+
$$
$$
+4\sum_{L\in {\cal S}\backslash {\cal S}_{\rm min}}\left[
\sum_{M\in {\cal S}\backslash {\cal S}_{\rm min}
:M<L}\mu^3(M)\left(1-p_M^{-1}\right)^2T{(M)}\sum_{I:M<I<L}\mu(I)\left(1-p_I^{-1}\right)T{(I)}\right]
{T{(L)}}^2+
$$
$$
+\sum_{L,M\in {\cal S}\backslash {\cal S}_{\rm
min}}\mu^2(L)\mu^2(M)\left(1-p_L^{-1}\right)\left(1-p_M^{-1}\right)T{(L)}T{(M)}{T{(\,{\rm
sup}\,(L,M))}}^2
$$
}
\end{lemma}

The proof is by application of lemmas \ref{prod_mat},
\ref{formulae}.

\section{Comparison with the Parisi case}

In the present section we compare the computed in the previous
section functionals of the introduced replica matrices with the
known results for the Parisi matrices. For the Parisi anzats the
corresponding tree looks as follows: the branching index $p_j$
depends on the integer $j$ (the level in the tree) for
$j=1,\dots,k$, $\mu(J)=m_{j}$, where
$$
m_{j}=\prod_{l=1}^{j} p_l,\qquad m_0=1.
$$
Values $T{(J)}$ of the function on a tree, in the case when the
function $T{(J)}$ depends only on the level $j$, are encoded into
$t^{(j)}$.

\begin{lemma}\label{to_Parisi_form}{\sl
Under the described above conditions the computed in the previous
section functionals of replica matrices reduce to the functionals
of the Parisi anzats as follows:
$$
J\in {\cal S}\backslash {\cal S}_{\rm min}\mapsto
j=1,\dots,k,\qquad \mu(J)\mapsto m_{j},\qquad T{(J)}\mapsto
t^{(j)},$$
$$
\sum_{J\in {\cal S}\backslash {\cal S}_{\rm min}}F(J)\mapsto
\sum_{j=1}^k{m_k\over m_j}f(j)
$$
In the sum above the coefficient ${m_k\over m_j}$ corresponds to
the multiplicity of vertices on the level $j$ of the tree. }
\end{lemma}

\noindent{\bf Example 1}\qquad Compute the functional
(\ref{sum_T_IJ}) of the form
$$
\sum_{I,J\in {\cal S}_{\rm min};I\ne J}\mu(I){\mu(J)}T{(\,{\rm
sup}\,(I,J))}=\sum_{J\in {\cal S}\backslash {\cal S}_{\rm
min}}T{(J)}\mu^2(J)\left(1-p_J^{-1}\right)
$$
We get for (\ref{sum_T_IJ}), normalized by the division by
$n=m_k=\mu(K)$ the following expression
$$
\sum_{j=1}^{k}t^{(j)}{m_k\over
m_j}\left[m_{j}-m_{j-1}\right]{m_j\over
m_k}=\sum_{j=1}^{k}t^{(j)}\left[m_{j}-m_{j-1}\right]
$$
where ${m_k\over m_j}$ is the degeneracy of $t^{(j)}$, and
${m_j\over m_k}$ is ${\mu(J)\over \mu(K)}$. This reduces to the
known answer of the Parisi replica symmetry breaking anzats
\be\label{P1} \sum_{j=1}^{k}t^{(j)}\left[m_{j}-m_{j-1}\right] \ee
where $m_0=1$ and $m_k=n$.

\bigskip

\noindent{\bf Example 2}\qquad Compute the functional
(\ref{ultra}) (normalized by division by $n=\mu(K)$) of the form
$$
{1\over\mu(K)}\sum_{L\in {\cal S}\backslash {\cal S}_{\rm
min}}\mu^3(L)\left(1-p_L^{-1}\right)\left(1-2p^{-1}_{L}\right)T_1{(L)}T_2{(L)}T_3{(L)}+
$$
$$
+{1\over\mu(K)}\sum_{L\in {\cal S}\backslash {\cal S}_{\rm
min}}\mu(L)\left(1-p_L^{-1}\right) \sum_{B\in {\cal S}\backslash
{\cal S}_{\rm min} :B<L}\mu^2(B)\left(1-p_B^{-1}\right)
$$
$$
\left[T_1{(L)}T_2{(L)}T_3{(B)}+T_2{(L)}T_3{(L)}T_1{(B)}+T_3{(L)}T_1{(L)}T_2{(B)}\right]
$$
We get
$$
{1\over m_k}\sum_{l=1}^k{m_k\over m_l}
m_{l}\left(m_l-m_{l-1}\right)\left(m_l-2m_{l-1}\right)t_1^{(l)}t_2^{(l)}t_3^{(l)}+
$$
$$
+{1\over m_k}\sum_{l=1}^{k}{m_k\over
m_l}\left(m_{l}-m_{l-1}\right) \sum_{b=1}^{l-1}{m_l\over
m_b}m_b\left(m_b-m_{b-1}\right)
\left[t_1^{(l)}t_2^{(l)}t_3^{(b)}+t_2^{(l)}t_3^{(l)}t_1^{(b)}+t_3^{(l)}t_1^{(l)}t_2^{(b)}\right]=
$$
$$
=\sum_{l=1}^k\left(m_l-m_{l-1}\right)\left(m_l-2m_{l-1}\right)t_1^{(l)}t_2^{(l)}t_3^{(l)}+
$$
$$
+\sum_{l=1}^{k}\left(m_{l}-m_{l-1}\right)
\sum_{b=1}^{l-1}\left(m_b-m_{b-1}\right)
\left[t_1^{(l)}t_2^{(l)}t_3^{(b)}+t_2^{(l)}t_3^{(l)}t_1^{(b)}+t_3^{(l)}t_1^{(l)}t_2^{(b)}\right]
$$
This again fits into the standard RSB picture, replica
computations of similar kind for the Parisi case were performed in
\cite{nature}.

\section{Appendix: trees, ultrametric spaces, PDO}

In the present section we, following papers \cite{ACHA},
\cite{Izv}, define a family of ultrametric spaces related to
trees. For discussion of relation between trees and ultrametric
spaces see also \cite{Serre}. An ultrametric space is a metric
space with the metric $|xy|$ (the distance between $x$ and $y$),
which satisfies the strong triangle inequality
$$
|ab|\le\hbox{ max }(|ac|,|cd|),\qquad \forall c
$$

We consider directed trees, i.e. trees with partial order, which
is a direction. A partially ordered set is called {\it directed}
(and the corresponding partial order --- a direction), if an
arbitrary finite subset has the unique supremum (remind that the
supremum of the subset of a partially ordered set is a minimal
element of the set, which is larger or equal to all elements of
the subset).

Consider an arbitrary tree ${\cal T}$ (finite or infinite), such
that the path in the tree between arbitrary two vertices is
finite, and the number of edges incident to each of the vertices
is finite. If a non--maximal vertex $I\in {\cal T}$ is incident to
$p_I+1$ edges, we will say that the branching index of $I$ is
$p_I$. If maximal index $I\in {\cal T}$ is incident to $p_I$
edges, we will say that the branching index of $I$ is $p_I$.
Equivalently, branching index of a vertex $I$ in directed tree is
the number of maximal elements, which less than $I$.

The absolute of a tree will be an ultrametric space (with respect
to the naturally defined metric). Consider two equivalent
definitions of the absolute of the tree.

The first definition is as follows. The infinitely continued path
with the beginning in vertex $I$ is a path with the beginning in
$I$, which is not a subset of a larger path with the beginning in
$I$. The space of infinitely continued paths in the directed tree
${\cal T}$, which begin in some vertex $R$ (that is, the root) is
called the absolute of the tree. Obviously the definition of the
absolute of the tree does not depend on the choice of $R$ (taking
any other vertex $A$ leads to an equivalent definition).

The equivalent definition of the absolute is as follows: the
absolute is the space of equivalence classes of infinitely
continued paths in the tree ${\cal T}$, such that any two paths in
one equivalence class coincide starting from some vertex (i.e. the
tails of the paths in one equivalence class are the same). If we
choose in each of the equivalence classes the paths, which begin
in vertex $R$, we will reproduce the first definition.

We consider trees with a partial order, where the partial order is
defined in the following way. Fix the vertex $R$ and the point
$\infty$ at the absolute. To fix the point $\infty$ at the
absolute means to fix the infinitely continued path $R\infty$ from
the vertex $R$ to $\infty$. The point $\infty$ we will call the
infinite point, or the infinity. We define the following natural
partial order on the set of vertices of the tree: $J>I$ if $J$
belongs to the path $I\infty$.

Consider the absolute with excluded infinite point, or
equivalently, the space of equivalence classes of decreasing paths
in ${\cal T}$. In the following we will call the absolute with
excluded infinite point the absolute. We denote the absolute of
the tree ${\cal T}$ by $X=X({\cal T})$ (note that we already
excluded the infinite point). Let us construct the ultrametric and
the measure on $X$.

For the points $x$, $y$ of the absolute there exists a unique path
$xy$ in the tree. The notation $xy$ should be understood in the
following way. Since the points $x$, $y$ of the absolute are
identified with the paths $Rx$ and $Ry$, the path $xy$ will be
contained in $Rx \bigcup Ry$. Then there exists a unique vertex
$A$ satisfying \be\label{A} Rx=RAx,\qquad Ry=RAy,\qquad Ax\bigcap
Ay=A \ee The notation $ABC$ means that $AC=AB\bigcup BC$. Then
$$
xy=Ax\bigcup Ay
$$
Define the vertex in ${\cal T}$, which is the supremum of $x$ and
$y$: \be\label{I} {\rm sup}(x,y)=xy\bigcap x\infty \bigcap y\infty
\ee Analogously, for vertices $A$, $B$ of the tree we define
\be\label{II} {\rm sup}(A,B)=AB\bigcap A\infty \bigcap B\infty \ee
as well as for $A\in{\cal T}$, $x\in X({\cal T})$
\be\label{III}{\rm sup}(A,x)=Ax\bigcap A\infty \bigcap x\infty\ee

Definitions (\ref{I}), (\ref{II}), (\ref{III}) make ${\cal T}$ and
${\cal T}\bigcup X({\cal T})$ the directed sets.

Put into correspondence to an edge in the tree the branching index
of the largest vertex of the edge (this definition is correct,
since any two vertices, connected by edge, are comparable). Then
the distance $|xy|$ is introduced as the product of branching
indices of edges in the directed path $RI$, $I={\rm sup}(x,y)$ in
the degrees $\pm 1$, where branching indices of increasing edges
are taken in the degree $+1$, and branching indices of decreasing
edges are taken in the degree $-1$. Here an edge is called
increasing, if the end of the edge is larger than the beginning,
and is called decreasing in the opposite case:
\be\label{distance3} |xy|=\prod_{j=0}^{N-1}
p^{\varepsilon_{I_{j}I_{j+1}}}_{I_{j}I_{j+1}},\qquad
I_0=R,\dots,I_{N}=I \ee where $\varepsilon_{I_{j}I_{j+1}}=1$ for
$I_{j}<I_{j+1}$, and $\varepsilon_{I_{j}I_{j+1}}=-1$ for
$I_{j}>I_{j+1}$.

\begin{lemma}\label{isultrametric}{\sl
The function $|xy|$ is an ultrametric (i.e. it is nonnegative,
equal to zero only for $x=y$, symmetric, and satisfies the strong
triangle inequality):
$$
|xy|\le \,{\rm max }\,(|xz|,|yz|),\qquad \forall z
$$}
\end{lemma}

To define the measure $\mu$, it is enough to define this measure
on the disks $I$, where disk $I$ is the set of all the infinitely
continued paths incident to the vertex $I$ which intersect the
path $I\infty$ only at the vertex $I$. Define the diameter $d_I$
of the disk as the supremum of the distance $|xy|$ between the
paths $Ix$ and $Iy$ in $I$. Then $I$ is the ball of radius $d_I$
with its center on any of $Ix\in I$.

\begin{definition}\label{measureeqradius}{\sl
The measure $\mu(I)$ of the disk $I$ is equal to the disk
diameter.}
\end{definition}

Since the disk $I$ contains $p_I$ maximal subdisks, which by
definitions of the ultrametric and the measure have the measure
$p_I^{-1}\mu(I)$, the measure $\mu$ is additive on disks. By
additivity  we can extend the measure on algebra generated by
disks ($\sigma$--additivity of the measure will follow from the
local compactness of the absolute, analogously to the case of the
Lebesgue measure). We denote $L^2(X,\mu)$ the space of the square
integrable (with respect to the defined measure) functions on the
absolute.

\begin{definition}\label{transprob}{\sl
We say that the operator in $L^2(X,\mu)$ of the form
\be\label{generator} T f(x)=\int T{({\rm
sup}(x,y))}(f(x)-f(y))d\mu(y) \ee where $T{(I)}$ is a function on
the tree ${\cal T}$, is an ultrametric pseudodifferential operator
(PDO).}
\end{definition}

\bigskip

\centerline{\bf Acknowledgements}

\medskip

The authors would like to thank G.Parisi and I.V.Volovich  for
fruitful discussions and valuable comments.  One of the authors
(A.Kh.) would like to thank S.Albeverio for fruitful discussions
and support of $p$--adic investigations. This paper has been
partly supported by EU-Network ''Quantum Probability and
Applications''. One of the authors (S.V.Kozyrev) has been partly
supported by the CRDF (grant UM1--2421--KV--02), by The Russian
Foundation for Basic Research (project 05-01-00884-a), by the
grant of the President of Russian Federation for the support of
scientific schools NSh 1542.2003.1, by the Program of the
Department of Mathematics of Russian Academy of Science ''Modern
problems of theoretical mathematics'', by the grant of The Swedish
Royal Academy of Sciences on collaboration with scientists of
former Soviet Union, and by the grants DFG Project 436 RUS
113/809/0-1 and RFFI 05-01-04002-NNIO-a.

\end{document}